# Performance Evaluation of Automated Multi-Service Deployment in Edge-Cloud Environments with the CODECO Toolkit

Georgios Koukis[a,1,*], Ioannis Dermentzis[a], Vassilis Tsaoussidis[a], Jan Lenke[b], Fabian Wölk[b], Daniel Uceda[c], Guillermo Sánchez[c], Miguel A. Puentes[d], Javier Serrano[d], Panagiotis Karamolegkos[e], Rute C. Sofia[f]

[a] *Democritus University of Thrace / Athena Research Center, Greece*
[b] *University of Göttingen, Germany*
[c] *Telefónica, Spain*
[d] *Universidad Politécnica de Madrid, Spain*
[e] *University of Piraeus Research Center, Greece*
[f] *fortiss GmbH, Germany*

## Abstract

Containerized microservices are widely adopted for latency-sensitive and compute-intensive applications, with Kubernetes (K8s) as the dominant orchestration platform. However, automating the deployment and management of multi-service applications remains challenging, particularly in heterogeneous Edge–Cloud environments. This paper evaluates the CODECO toolkit, an open-source framework designed to enhance container orchestration across distributed infrastructures. We compare CODECO with baseline K8s workflows using three key performance indicators: deployment time, level of manual intervention, and runtime performance with resource utilization. Experiments across diverse hardware platforms (ARM, AMD, RPi) and K8s distributions, including lightweight variants such as k3s, demonstrate that CODECO substantially reduces manual effort while maintaining competitive performance and acceptable overhead. These results validate CODECO as an effective solution for Edge-Cloud orchestration and highlight its potential to improve the flexibility and intelligence of K8s-based deployments.

## 1. Introduction

Edge–Cloud computing has emerged as a key architectural paradigm, distributing computation, storage, data handling, and intelligence across the Edge–Cloud–IoT (CEI) continuum. It has fundamentally reshaped how digital services are designed and operated, shifting from centralized cloud architectures toward decentralized, distributed systems. This transition has accelerated the adoption of microservice-based architectures and containerization technologies (e.g., Docker), enabling applications to run across heterogeneous resources, including cloud data centers, edge servers, and resource-constrained IoT devices. As a result, Edge–Cloud computing has become a standard across multiple application domains, such as smart manufacturing, Industrial Internet of Things (IIoT) [1], intelligent transportation systems [2], Content Delivery Networks (CDN) [3], energy systems [4] and smart cities [5], where low latency, data locality, resilience and scalability are critical requirements.

In this context, Kubernetes (K8s) has become the de facto orchestrator for operating microservices, offering automation features such as multi-node scheduling, load balancing, autoscaling and service discovery. However, the distributed nature of the CEI continuum introduces novel operational challenges, particularly in managing dynamic, heterogeneous workloads and infrastructures [6, 7], maintain performance and resource footprint [8, 9] or incorporating intelligence and context-awareness into decision making [10, 11]. Practical deployments mix hardware (ARM/x86), cluster topologies, resource profiles and varying network conditions, requiring intelligent and secure mechanisms (e.g., leverage ML/AI) and data workflows with continuous monitoring and real-time adaptability features.

To address these needs, the open-source[2] CODECO (Cognitive Decentralised Edge Cloud Orchestration) project [12] introduces a modular, decentralized toolkit for orchestration across far Edge-Cloud. It streamlines containerized service deployment and lifecycle management, enabling cognitive, cross-layer orchestration across diverse data, compute, energy and networking, while supporting intelligent resource utilization and dynamic offloading. In addition, the CODECO Experimentation Framework (CODEF) [13] supports reproducible experimentation via declarative experiment specifications and parameterized cluster/application setup across heterogeneous environments, thereby minimizing human intervention and enabling automation.

*Corresponding author.
*Email address:* gkoukis@ee.duth.gr,
ORCID=0009-0005-0116-7296 (Georgios Koukis)

[1] This work has been funded by The European Commission in the context of the Horizon Europe CODECO project under grant number 101092696 and by the Swiss State Secretariat for Education, Research and Innovation (SERI) under contract number 23.00028.

[2] https://gitlab.eclipse.org/eclipse-research-labs/codeco-project and https://projects.eclipse.org/projects/technology.kudeco

This study evaluates the performance of CODECO in multi-service deployment across a variety of workload scenarios and real-world infrastructures provided by CODECO partners. This includes on-premises or shared cloud servers, lightweight edge clusters based on Raspberry Pi and virtualized environments. Deployments utilize both vanilla K8s and lightweight distributions such as k3s, covering a spectrum of Edge–Cloud configurations. Our evaluation considers multiple scenarios, from idle cluster state to stress and scaling workloads, using both benchmark and dummy applications, including the widely adopted Bookinfo microservice suite.

In summary, the main contributions of this work are as follows:

- A quantitative evaluation of the CODECO open-source toolkit assesses the automation and orchestration performance of multi-service deployments using key performance indicators (KPIs) related to deployment, performance, and manual intervention.
- A comprehensive, multi-infrastructure experimental study is conducted across diverse hardware configurations and K8s distributions, spanning experimental testbeds using the novel experimentation framework CODEF and real-world Edge-Cloud CODECO use cases.
- Performance evaluation of the CODECO open-source toolkit is conducted using three KPIs, assessing cluster lifecycle and resource/energy footprint across multiple applications and workload scenarios with diverse scalability and stress characteristics.
- The experimentation results are made openly available through the project repository to enable transparency, reproducibility and further research.
- A reusable experimentation methodology compliant with the Internet Engineering Task Force (IETF) Benchmarking (BMWG) working group guidelines is incorporated via the CODEF framework to support benchmarking and reproducibility.

The remainder of this paper is organized as follows. Section 2 discusses related works and initiatives on Edge–Cloud orchestration and automation frameworks, while Section 3 presents CODECO, its architectural components, model and use cases, as well as the experimentation framework CODEF. Section 4 details the experimental setup and methodology, while Section 5 analyzes the obtained results. Finally, Section 6 concludes the paper and outlines our future directions.

## 2. Related Work

While this work focuses on evaluating the CODECO open-source Basic Operation toolkit under a single-cluster deployment model using specific performance parameters, rather than performing direct comparisons with alternative frameworks, it is useful to position our contribution within the broader research landscape by highlighting related works and EU-funded initiatives on AI-driven Edge–Cloud orchestration, serverless execution, and autonomous resource management frameworks.

Existing literature extensively explores AI-driven resource management and dynamic workload scaling across the edge–cloud continuum, with the aim of optimizing execution latency, resource utilization, and energy efficiency [14, 15]. Several studies investigate intelligent service management over heterogeneous infrastructures, including automated workload offloading, serverless execution models, and adaptive resource provisioning [16, 17].

More recent work combines serverless architectures with machine learning (ML)-based control mechanisms to enable automated deployment and dynamic scaling of functions in distributed environments, supporting elastic execution while addressing performance and efficiency challenges in the edge–cloud continuum [18, 19]. In addition, multi-cloud and far-edge orchestration has been explored through AI-driven coordination and adaptive scheduling techniques that improve distributed deployment efficiency and resource management across heterogeneous infrastructures [20, 21]. Finally, hierarchical and multi-agent AI architectures have been proposed for autonomous management across the device–edge–cloud continuum, enabling QoS-aware deployment and orchestration of distributed applications over heterogeneous resources, while improving system reliability and energy efficiency through closed-loop learning and control mechanisms [22, 23].

Complementary to the academic body of work, several large-scale initiatives address similar challenges from a systems and platform perspective. ENACT [24] applies AI-driven techniques for adaptive resource management and dynamic scaling of data-intensive workloads across the edge–cloud continuum. COGNIT [25] focuses on intelligent service management over heterogeneous infrastructures, supporting serverless execution and workload offloading. EDGELESS [26] leverages serverless paradigms and ML-based control for automated deployment and reconfiguration, enabling elastic scaling while maintaining integration with cloud services and promoting sustainability through performance-aware resource management. NEPHELE [27] addresses multi-cloud and far-edge orchestration through multi-cluster coordination, adaptive scaling and integrated monitoring, demonstrating efficient management of service deployment and distributed execution overheads. ML-SysOps [28] adopts a hierarchical multi-agent AI architecture for autonomous management across the device–edge–cloud continuum, supporting QoS-aware orchestration and energy-efficient operation over heterogeneous resources.

In contrast to these approaches, which primarily emphasize multi-cloud coordination, far-edge deployment, or AI-driven global optimization, CODECO focuses on providing a lightweight, semantically rich abstraction model and orchestration framework for Edge-Cloud nativeu applications. In this work, we specifically evaluate the CODECO Basic Operation toolkit in a controlled single-cluster setting, aiming to characterize its operational behavior through well-defined performance parameters and establish a baseline for future extensions toward multi-cluster and edge-aware deployments.

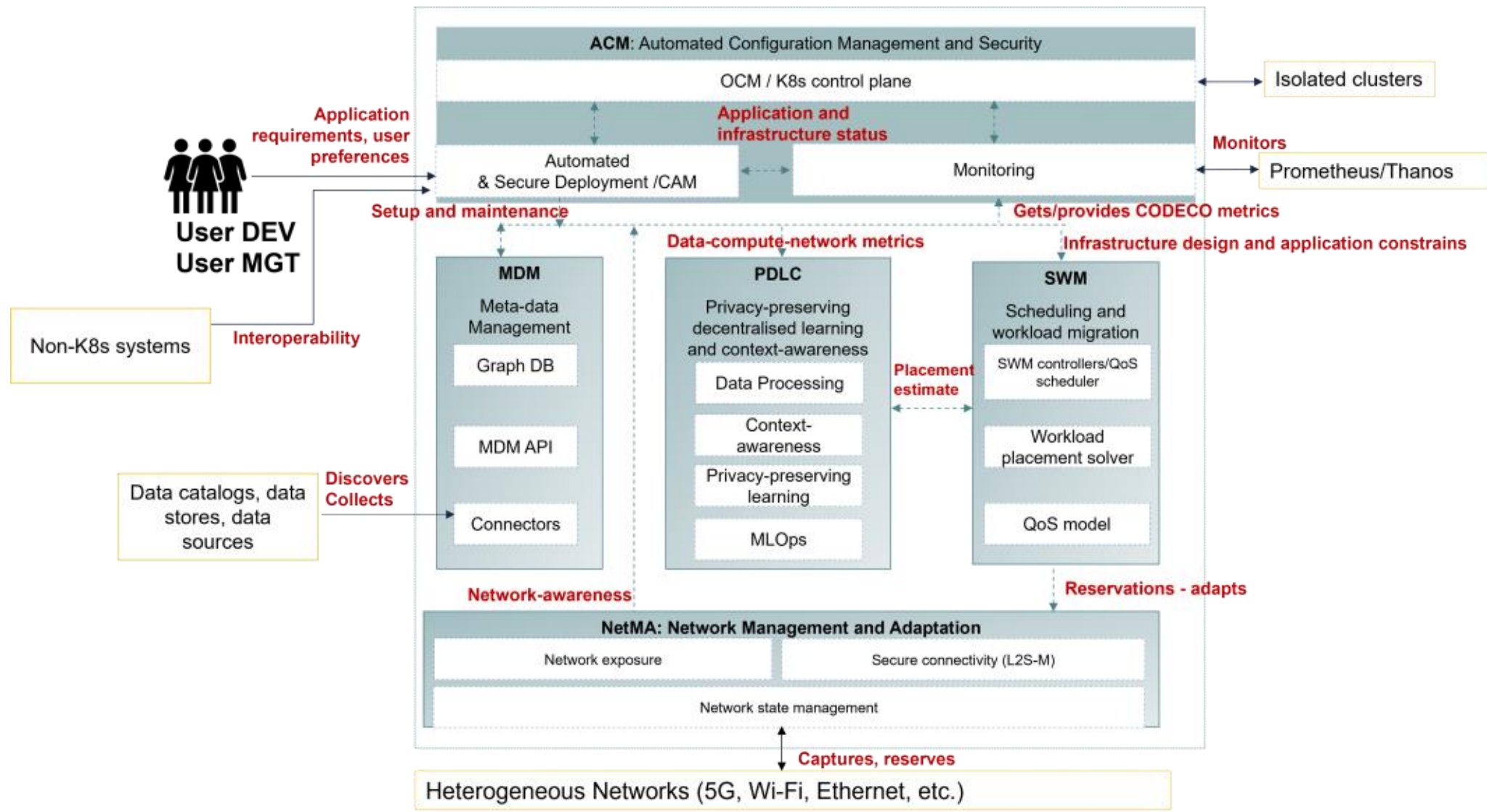

Figure 1: CODECO system model and components.

## 3. The CODECO Framework

CODECO is a novel open-source orchestration framework that extends K8s native capabilities with cognitive and policy-driven control for distributed Edge-Cloud orchestration of microservices. It interoperates with standard K8s but extends it with cross-layer abstractions (compute, network, data) and ML-/rule-based decisions. The goal is to deliver an orchestration plane that continuously adapts to application requirements, end-user intent and infrastructure conditions, ranging from data characteristics and resource footprint to network dynamics and user-level objectives, without requiring manual actions. Its components are illustrated in Fig. 1. CODECO consists of five collaborative components, briefly described below. A detailed description of the architecture, concepts and components can be found in [12].

### 3.1. Automated Configuration Management (ACM)

ACM is the user entry point of CODECO and managed operations across far Edge-Cloud for end-to-end configuration. It translates developer/operator inputs e.g., application/user requirements across networking, compute and data, into deployable specifications, installs CODECO components and triggers integration with external systems when required. ACM is co-located within the K8s control plane, managing cluster-level tasks and exposing a unified control interface to the user. Observability is integrated via Prometheus, currently focusing on compute, network, energy, greenness and data telemetry.

### 3.2. Scheduling and Workload Migration (SWM)

SWM performs coordinated placement and re-placement of microservices produced at setup time by ACM. It replaces the default K8s scheduler which makes point-in-time, cluster-local assignments rather than predictive orchestration, planning ahead or optimizing across heterogeneous clusters. SWM formulates placement as a constrained decision problem over the considered metrics (QoS, resources, network, greenness), leveraging a solver to match user/application intents and perform migration actions based on the recommendations acquired from the PDLC component.

### 3.3. Privacy-preserving Decentralized Learning and Context-awareness (PDLC)

PDLC provides orchestration with context derived from telemetry exposed by ACM, NetMA and MDM. It computes node-level weights aligned with application/user-selected profiles (e.g., resilience, energy efficiency) using a mix of ML, rule-based logic and decentralized, privacy-preserving learning. As the core brain of CODECO, it turns each cluster node into a context-aware resource, enabling placement decisions that adapt to changing conditions.

### 3.4. Network Management and Adaptation (NetMA)

NetMA contributes to the network awareness and connectivity control of CODECO, exposing network attributes relevant to placement quality (e.g., latency, bandwidth) and establishing secure pod-to-pod paths. For programmable data paths, NetMA integrates SDN–K8s interaction (via the L2S-M component) and maintains a view of network state.

### 3.5. Metadata Manager (MDM)

MDM provides data observability to the rest of the CODECO components, capturing and correlating metadata across application, system and network layers at key orchestration points. It enables data traceability and performance impact analysis, complementing ACM and NetMA's system and network monitoring.

Table 1: Indicative CAM fields (YAML keys) and their intent.

| Field / YAML key | Description |
|---|---|
| performanceProfile | Optimization goal (e.g., energy efficiency vs. performance/cost). Used to steer scheduling/placement decisions (e.g., prioritize greener/lower-energy nodes when set to Greenness). |
| appEnergyLimit | Application-level energy budget (provided as a string, e.g., "20"). Enables energy-aware scheduling decisions based on an upper bound of energy consumption. |
| appFailureTolerance | Acceptable failure tolerance level for the application. Can be an empty string ("") (i.e., no explicit tolerance provided). |
| schedulerName | Scheduler selection for the workload (e.g., qos-scheduler instead of K8s' default-scheduler). |
| serviceChannels | Declares network channels between microservices with QoS requirements/guarantees (i.e., application-intent networking beyond vanilla K8s. |
| serviceClass | Network traffic/service class (e.g., ASSURED, BESTEFFORT). Used to request bandwidth/latency guarantees. |
| bandwidth | Requested/guaranteed bandwidth between services (e.g., "5M" for 5 Mbps). |
| maxDelay | Maximum allowed end-to-end delay/latency between services (e.g., "10ms" as specified). |
| complianceClass | Compliance requirements for the application (e.g., data handling constraints, placement/processing). |
| qosClass | Service tier for prioritization (e.g., Gold > Silver > Bronze). |
| securityClass | Security requirements for the application (e.g., baseline security level expected from hosting environment/policies). |

### 3.6. *CODECO Application Model (CAM)*

The CODECO Application Model (CAM)[3] is a YAML-based, declarative manifest created by the user via the ACM interface during application deployment setup. The CAM describes the service(s) that make up an application and the requirements under which they should run (e.g., QoS, network, energy) that ACM assembles and validates (rf. to Table 1). The finalized CAM informs cluster sizing and resource reservations, transferred to the CODECO components to drive placement, migration and end-to-end lifecycle management.

### 3.7. *CODECO Experimentation Framework (CODEF)*

CODEF (rf. to Fig. 2) is an open-source framework developed to accelerate experimentation in containerized Edge–Cloud environments. Its microservice-oriented, modular design promotes flexibility, extensibility and the seamless incorporation of new technologies. CODEF provides abstractions that cover the full experimental workflow, i.e., from node allocation and cluster/application deployment to declarative experiment specification (including fine-grained parameterization e.g., K8s flavor, CNI, security, application, scope) and end-to-end automation from system setup through results visualization. It offers a comprehensive, adaptable approach to manage heterogeneous resources across diverse infrastructures. By building on and extending well-established tools (e.g., Ansible, kubeadm, Bash, SSH), CODEF ensures automation, interoperability and reproducibility for complex experimental scenarios. Details on the architecture, proof-of-concept implementations, experimentation and validation can be found in [13].

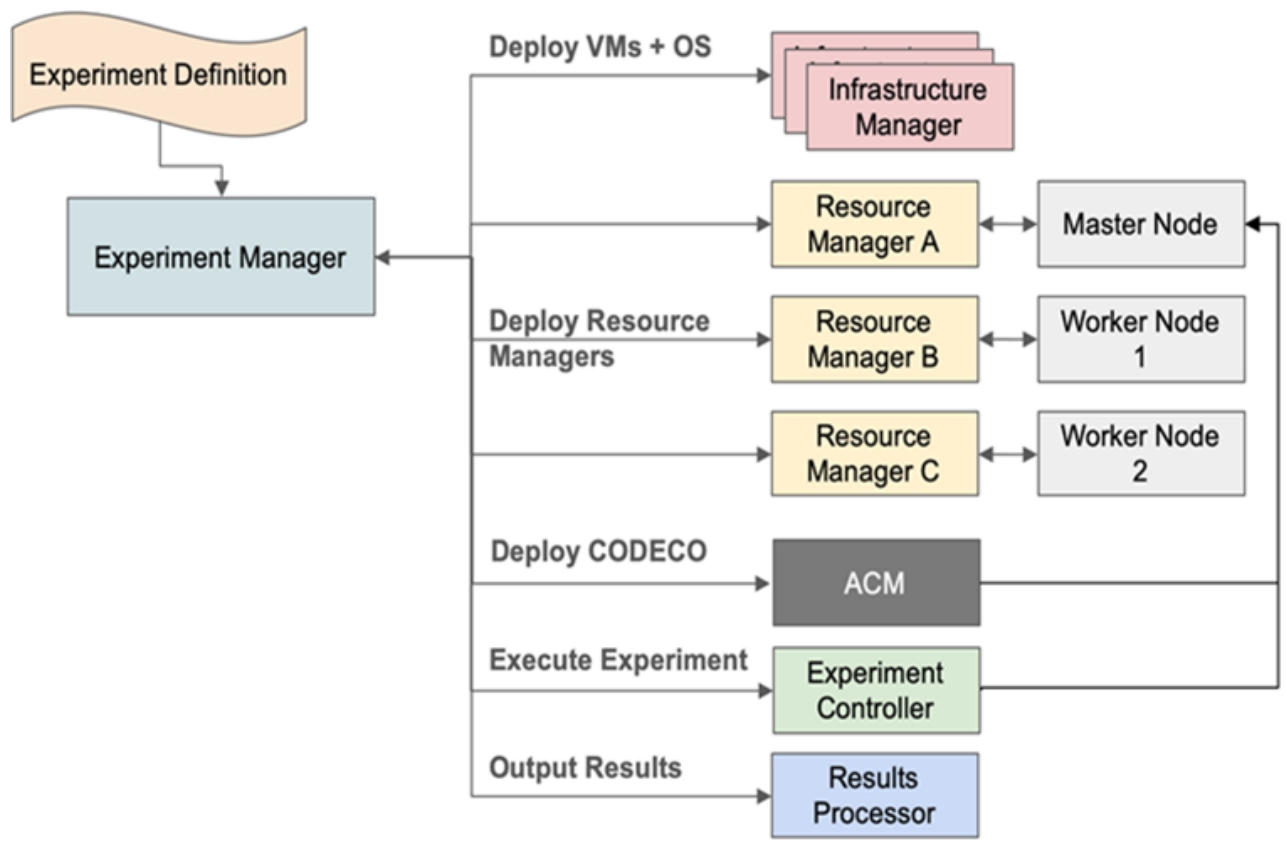

Figure 2: CODEF high-level architecture.

### 3.8. *CODECO Use Cases and Infrastructures*

This section outlines four selected use cases (UCs) which are being used in the Horizon Europe CODECO project, to assess the peformance and orchestration support that CODECO can provide in different vertical domains. Each UC leverages the CODECO toolkit to evaluate its automation, orchestration and performance gains, in addition to adaptation capabilities under heterogeneous and domain-specific conditions. The selected UCs span a range of verticals, including smart-city and buildings, decentralized energy management and industrial/robotics automation, covering both experimental testbeds and real-world deployments. Detailed specifications on architecture and objectives of each UC are available in [12] and [29].

#### 3.8.1. *UC1: Smart Monitoring of the Public Infrastructure (UGOE)*

Deployed in the smart-city of Göttingen (Germany), UC1 enhances urban mobility and safety by deploying real-time traffic and pedestrian analytics across the edge–cloud setup, in-cluding thermal cameras/LiDARs on edge nodes (K8s workers) with a cloud-hosted control plane. CODECO orchestrates low-latency processing and adaptive (re)placement of services, securing data flows across heterogeneous resources. The goal is to include congestion reduction, pedestrian safety and planning support via continuous dashboards, with GDPR-compliant and privacy-preserving data handling. Evaluation focuses on deployability and runtime efficiency, demonstrating that CODECO can sustain real-time monitoring under the practical constraints of a smart-city-scale environment.

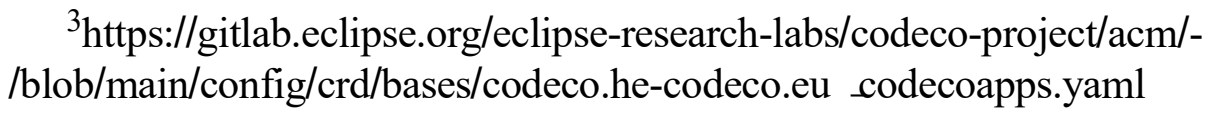

[3]https://gitlab.eclipse.org/eclipse-research-labs/codeco-project/acm/-/blob/main/config/crd/bases/codeco.he-codeco.eu _codecoapps.yaml

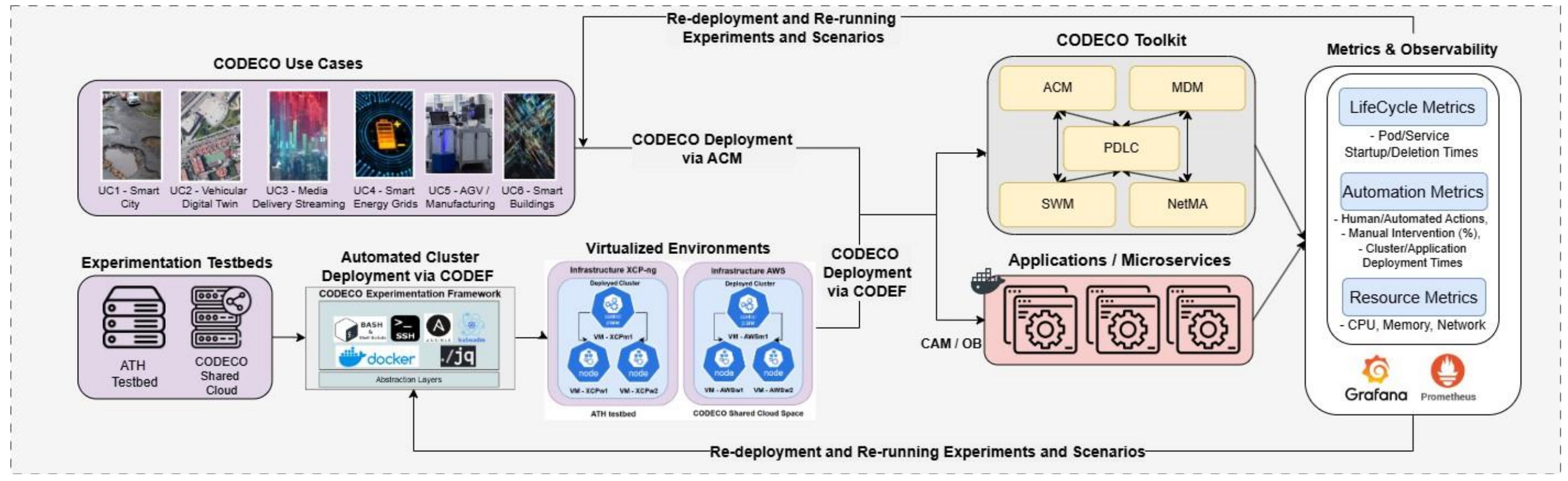

Figure 3: Experimentation methodology workflow.

#### 3.8.2. UC3: MDS across Decentralized Edge-Cloud (TID)

UC3 is deployed across Telefo´nica's virtual test environment (Spain) and ICOM's edge cluster (Greece), optimizing end-to-end media delivery across multi-domain Edge–Cloud infrastructures. It orchestrates compute placement and network paths, integrating the Media Delivery System with CODECO and NetMA to expose ALTO-style[4] topology and cost maps, enabling per-request selection of the optimal route. By aligning client demand with real-time network conditions it minimizes latency and hop count while improving QoE and resource efficiency. The UC focuses on evaluating accurate topology reflection and frequent network updates, selection of paths/servers to maximize QoE under network constraints, while ensuring secure exposure (via L2S-M).

#### 3.8.3. UC4: Demand Side Management in Decentralized Grids (UPM)

Deployed across Universidad Polite´cnica de Madrid campuses and selected landmark buildings (Spain), UC4 implements a decentralized demand-response system for energy optimization and building decarbonization. Leveraging CODECO's edge–cloud orchestration, energy-related IoT data are processed near the source, while aggregation, forecasting and optimization occur in the cloud to enable coordinated control under latency and reliability constraints. UC4 enhances energy efficiency, sustainability and resilience through cluster-level coordination and community services such as flexibility markets and EV charging. The UC focuses on evaluating dynamic energy-cluster formation, reduction in grid consumption and $CO_2$ emissions and sub-minute function deployment.

#### 3.8.4. UC5: Wireless AGV Control for Flexible Factories (FOR)

UC5 is deployed in the fortiss IIoT Lab (Munich), relying on CODECO to orchestrate the workloads deployed across a fleet of Autonomous Mobile Robots (AMR) interconnected via Wi-Fi. AMRs act as K8s workers running containerized microservices with CODECO managing deployment, placement and migration across compute and network layers to handle interference and intermittent links. The UC evaluates three configurations: single-cluster with static control, single-cluster with mobile control and federated clusters—to quantify end-to-end task completion time, recovery after wireless disruptions and energy per task. It targets lower task latency and collision rates, improved energy/network efficiency and robust operation under mobility, with scalability and high availability as key non-functional goals.

## 4. Experimentation Setup, Methodology and Metrics

This section provides an overview of the experimentation approach (rf. to Fig. 3) considered in this work, including the methodology, use case technical specifications, the considered Key Performance Indicators (KPIs) and associated metrics, as well as the evaluated workload scenarios. All experiments are fully reproducible with the complete implementation, configuration scripts and collected results available as open source under the project repository [30].

### 4.1. Methodology

The experimentation methodology follows recognized standards, specifically integrating guidelines from the IETF BMWG drafts e.g., the Telco-Cloud Benchmarking [31] and related Network Function Virtualisation (NFV) testing frameworks [32]. This work adopts suggestions on workload profiles and traffic models, metrics taxonomy (availability, resource efficiency) and test orchestration to ensure repeatability.

[4]https://datatracker.ietf.org/wg/alto/

Each scenario is executed multiple times to ensure accuracy and statistical robustness. We include (when applicable) *MIN*, *MAX*, *AVG*, *P95* and *STDIV* to showcase variability across experiment stages. We capture deployments with timestamps at key phases (e.g., Idle, pods Ready, Delete) and capture resource usage via cluster telemetry, utilizing Prometheus for metrics scraping, complemented by script logging. Our experiments run CODECO against plain K8s, which is considered our baseline. Across the evaluated hardware-heterogeneous Edge–Cloud use cases, scenarios and applications, we focus on defined KPIs and collect three classes of metrics:

- **Lifecycle:** measure the startup and deletion times of pods and services.
- **Automation:** quantify the required actions and the Manual Intervention Percentage.
- **Resource-footprint:** capture CPU, memory and energy utilization during cluster operations.

### 4.2. Key Performance Indicators and Metrics

To quantify the performance of the CODECO toolkit and the efficiency of automated multi-service deployment within single-cluster edge–cloud environments, this work focuses on a set of KPIs and corresponding experimentation metrics:

- **KPI1**: Time taken to set up and configure multiple clusters and services based on scalable single clusters.
- **KPI2**: Percentage of manual intervention required against K8s when setting up clusters and services.
- **KPI3**: Overall resource footprint (CPU, RAM and energy).

CODECO performance across the defined KPIs is assessed by collecting the following metrics, with vanilla K8s (i.e., K8s without CODECO components) serving as the baseline for comparison.

#### 4.2.1. Percentage of Manual Intervention

End-to-end deployment in Edge-Cloud environments involves a sequence of high-level manual actions across multiple stages, including:

1. **VM and OS provisioning**, including resource allocation, OS bootstrapping, SSH access setup, host configuration and base package installation.
2. **K8s installation**, covering container runtime, K8s setup and networking configuration (CNI installation).
3. **Service deployment**, requiring preparation of dependencies (e.g., Helm, yq, cert manager) and deploying the associated software components. For instance, in the CODECO case, this includes the deployment of ACM, NetMA, MDM, PDLC and SWM.

In summary, the Manual Intervention Percentage (MIP) is computed by identifying all major deployment and installation phases and quantifying the proportion of user-defined manual actions required in the CODECO and CODEF environment ($C_a$) relative to a fully manual K8s baseline ($K_a$), as defined in Eq. (1), where $C_a$ denotes the number of manual actions required in the CODECO/CODEF environment, and $K_a$ represents the corresponding number of manual actions in the fully manual K8s baseline.

$$\text{MIP} = \frac{C_a}{K_a} \times 100\% \tag{1}$$

Therefore, a lower MIP indicates a higher degree of automation. For the cluster deployment and K8s installation phases, CODEF-based automation (Ansible/Bash) is employed, providing end-to-end provisioning logic from host preparation and cluster bootstrap to application deployment across heterogeneous environments. In contrast, service installation relies on the CODECO *post_deploy.sh* script[5] , which installs and configures the complete CODECO toolkit via ACM.

#### 4.2.2. Cluster Deployment and Installation Times

This metric captures the time required for cluster deployment across distinct stages. For deployments performed with CODEF, we explicitly measure node discovery (*ND*), operating system installation ($OS_I$) and K8s installation ($K8S_I$), as CODEF manages the full node lifecycle, prior to cluster formation. In contrast, emulation environments based on KinD[6] assume pre-configured container-based nodes and therefore include only the cluster deployment and application installation stages. Each stage is measured based on explicit start and completion events to capture the deployment overhead across different cluster sizes.

#### 4.2.3. Pod Startup and Deletion Times

To assess pod initialization overhead, we measure the duration between pod creation requests (e.g., starting with the *kubectl apply* command) and the moment pods transition into *Ready* state as reported by K8s and Unix timestamping. It captures the total latency of pod initialization process, including image pulling, container creation, network setup and readiness probe completion, reflecting how quickly a system can bring new workloads online. In the context of CODECO, it captures the interval from ACM submitting an *ApplicationGroup* to all pods reaching *Ready* state. The interval includes SWM scheduling using PDLC context (e.g., bandwidth, energy, resource load), NetMA bringing network paths up and MDM supporting observability. Therefore, this metric indicates how CODECO's end-to-end coordination manages initialization latency across heterogeneous Edge–Cloud nodes.

[5]https://gitlab.eclipse.org/eclipse-research-labs/codeco-project/acm/-/blob/main/scripts/post_deploy.sh

[6]https://kind.io

Table 2: List of use case infrastructures and specification details.

| Name | Partner | Domain | Infrastructure Specifications | Kubernetes Environment | Deploy Method |
|---|---|---|---|---|---|
| INF1 | UPRC | Local deployment | Intel Core i9-10900K CPU @ 3.70 GHz, 10 cores / 20 threads (x86_64), 64 GB DDR4 RAM, NVIDIA GeForce GTX 1660 SUPER GPU, 1 TB SSD storage, running Ubuntu 22.04 with Linux kernel 6.8.0. | KinD | post_deploy |
| INF2 | UGOE | Smart City Monitoring (UC1) | *Control plane:* OpenStack VM, 16 GB RAM, 148 GB storage, 4 cores (AMD EPYC, @ 3.245 GHz). *Workers:* NVIDIA Jetson AGX Orin, 64 GB RAM, 12 vCores (ARMv8), 520 GB storage. | Vanilla K8s (VM-based) | post_deploy |
| INF3 | TID | Media Delivery Streaming (UC3) | *Control plane:* Proxmox VM (16 GB RAM, 32 GB storage, 8 cores, QEMU). *Workers (2):* Raspberry Pi 5, Broadcom BCM2712 quad-core 64-bit Arm Cortex-A76 @ 2.4 GHz, 4 GB RAM, 32 GB, Raspberry Pi OS. | k3s | post_deploy |
| INF4 | UPM | Smart Energy Grids (UC4) | *Control plane:* Intel Core i9-10900 @ 2.80 GHz, 10 cores/20 threads (KVM); 32 GB DDR4 DIMM @ 2933 MT/s; 1 TB NVMe SSD. *Worker 1:* Intel Core i7-7700K @ 4.20 GHz; 4 cores/8 threads (KVM); 16 GB DDR4 DIMM @ 2400 MT/s; 256 GB SATA SSD + 1 TB HDD. *Worker 2:* Intel Core i7-860 @ 2.80 GHz; 4 cores/8 threads (VT-x); 8 GB DDR3 DIMM @ 1333 MT/s; 2×500 GB HDD (SATA). | k3s | post_deploy |
| INF5 | FOR | Wireless AGV / Manufacturing (UC5) | *Control plane:* Intel(R) Core(TM) i7-8550U CPU @ 1.80 GHz; 16 GB DDR4. *Workers (7):* Raspberry Pi 4 (ARM64). | k3s-edge | post_deploy |
| INF6 | ATH | Experimentation CODECO Shared Cloud Space (AWS) | Per node specifications (3 nodes): Intel Xeon Platinum 8259CL @ 2.50 GHz, 8 physical cores, 16 threads (KVM); 64 GB DDR4 DIMM @ 3200 MT/s; 97 GB NVMe SSD. | Vanilla K8s (VM-based) | CODEF |

#### 4.2.4. Service Startup and Deletion Times

To evaluate service instantiation overhead, we measure the interval from the service deployment (i.e., when a K8s Service object and associated pods are created) until the service becomes fully operational and accessible. Conversely, the service deletion time measures the time taken for a service and all associated resources (e.g., pods, deployments, endpoints) to be terminated and removed from the cluster. This cluster lifecycle metric captures CODECO's end-to-end service management from ACM/SWM-driven instantiation and placement based on PDLC recommendations, to endpoint availability through NetMA channel establishment and conversely, from deletion request to complete teardown.

#### 4.2.5. Resource Consumption

This metric captures the overall resource efficiency and energy footprint of CODECO and deployed applications. Specifically, we measure CPU utilization (in %), memory consumption (in GB) and power usage (in W and kWh/day) during idle phase. Resource metrics are collected on control plane and worker nodes, reporting both per-node CPU and memory, but also cluster/system-level energy. The data are collected via Prometheus exporters or *top* commands (for system-level metrics), comparing CODECO with baseline K8s or KinD, highlighting how CODECO's scheduling, context- and network-awareness impact resource utilization.

Estimating energy consumption in environments which combine VM-based infrastructures (e.g., CODEF) and containerized local clusters (e.g., KinD) poses significant challenges. Tools such as Kepler[7] leverage eBPF and hardware counters to estimate energy at the container and process level, however, their accuracy is limited in virtualized environments, thus, reliance on pre-trained power models may be problematic, as discussed in the IETF GREEN and BMWG Working Groups [33, 31]. Given these constraints, we adopt an established linear power model [34]:

$$P = P_{idle} + (P_{max} - P_{idle}) \times U_{cpu} + P_{mem} \qquad (2)$$

where $P$ is the instantaneous power draw ($W$), $P_{idle}$ and $P_{max}$ represent the idle and peak power consumption, derived from hardware specifications, $U_{cpu}$ is the measured CPU utilization (0-1), and $P_{mem}$ is the memory power proportional to usage

[7]https://sustainable-computing.io/

(≈0.2 W/GB). Daily energy consumption ($E$) is then computed as $E = P \times 24$ h.

### 4.3. Evaluation Use Cases and System Specs

This section describes the underlying infrastructures of the CODECO technical use cases. We evaluate the deployed scenarios across six heterogeneous environments (INF1 to INF6), including two experimental and four on-premise real-world testbeds, covering diverse hardware (ARM/x86, edge devices to servers), network conditions and K8s flavors (vanilla, k3s). In addition, to evaluate scalability-related KPIs, we also run a subset of experiments on a local KinD cluster hosted on a personal computer (in INF1). Detailed infrastructure specifications and deployment methods are summarized in Table 2.

### 4.4. Experimental Applications

This section defines the application workloads and tools used to evaluate CODECO's deployment and orchestration performance in single-cluster environments. We execute a range of applications, starting from simple batch pod deployments, extending to a dummy application and finally a realistic microservice benchmark suite (Bookinfo), in order to study initialization, scaling and teardown behavior under heterogeneous nodes and varying workload profiles. In addition to standard *kubectl*-based deployments, all applications are also deployed through the CAM to demonstrate CODECO's orchestration capabilities.

#### 4.4.1. Application 1: Pod readiness and Deletion per pod batch

In our experiments, we launch batches of 10–150 pods on cluster workers using a *k8s.gcr.io/pause:3.1*) minimal image. The pods execute no application workload; for each batch we record the elapsed time until pods reach the *Ready* condition and report the batch-average readiness and deletion time.

#### 4.4.2. Application 2: Dummy Frontend-Backend

We deploy a lightweight backend and a frontend service[8], packaged as independent containers and establishes a bidirectional service channel between them to simulate a typical microservice workflow. In particular, the backend provides a simple HTTP endpoint (e.g., */api/hello*) that returns greeting text, while the frontend issues requests to this endpoint and displays or processes the responses.

#### 4.4.3. Application 3: Bookinfo

Bookinfo[9] is a microservice-based benchmark application that represents a simple online book catalog. It consists of independent services (e.g., *productpage*, *details*, *reviews*, *ratings*) that interact through HTTP, simulating controlled traffic and service behavior. In our experiments, we deploy the full Bookinfo stack to quantify startup latency and orchestration efficiency across heterogeneous nodes.

[8] https://gitlab.eclipse.org/eclipse-research-labs/codeco-project/acm/-/blob/main/config/samples/codeco_v1alpha1_codecoapp_ver3.yaml

[9] https://istio.io/latest/docs/examples/bookinfo/

## 5. Experimental Results

### 5.1. Percentage of Manual Intervention

Table 3 summarizes the user effort (as actions) required to: (i) provision a cluster (VM and OS), (ii) install K8s and (iii) deploy the CODECO toolkit, comparing the baseline K8s (manual) approach and the CODECO-driven (automated) procedures, complemented by CODEF processes, specifically for the first two phases. Via the Ansible playbooks of CODEF we can enumerate the required actions of these phases as per playbook tasks.

Table 3: Manual intervention against baseline K8s.

| Phase | Method | Auto Level | Actions | MIP (%) |
|---|---|---|---|---|
| Cluster Deployment | Baseline K8s | None | 19 | 25.3 |
| | CODECO | Full | 3 | |
| K8s Installation | Baseline K8s | None | 42 | 9.52 |
| | CODECO | Full | 4 | |
| Service Deployment | Baseline K8s | None | 11 | 18.18 |
| | CODECO | Full | 2 | |

Regarding cluster deployment (phase I), the baseline manual approach requires a total of 19 discrete actions, including VM instantiation on the target infrastructure, OS image selection, SSH key generation and distribution, firewall rule setup and more. CODEF reduces the user-defined actions to 3 (i.e., MIP=25·3%), including the creation of the configuration file (with the infrastructure manager used, user credentials and nodes' specifications), the declaration of the above file in the deployment script and the actual execution process.

In terms of K8s installation (phase II), actions such as Docker/containerd runtime installation, kubeadm initialization, certificate generation, K8s distribution and CNI plugin deployment, worker node joining, and verification steps must be performed, a total of 42 actions. CODEF automates the procedure by requiring the user to select the required K8s distribution, CNI plugin, any other network parameter, and finally run the script, i.e., 4 actions which are equal to 9.52% MIP.

Focusing on service deployment (phase III), since the functionalities provided by CODECO extend beyond native K8s capabilities, we make the following assumptions. We quantify the manual actions that would be required to replicate CODECO's functionality in vanilla K8s, assuming the equivalent components were available but not automated. Under this scenario, the manual service deployment process includes the retrieval of cluster information and credentials, installation of prerequisites (e.g., Prometheus and Kepler), cloning of component repositories, their deployment (ACM, NetMA, MDM, PDLC, SWM), validation, and verification of end-to-end functionality, a total of 11 discrete actions. In contrast, CODECO automates this entire workflow through the execution of the *post_deploy.sh* and validation scripts. Consequently, the resulting MIP is calculated at 18·18%.

### 5.2. Cluster Deployment and Installation Times

Fig. 4 represents the mean deployment times as the number of nodes scales, obtained from (i) CODEF and (ii) a KinD-based environment.

Fig. 4(a) breaks down CODEF deployment time into Node Discovery (*ND*), OS installation ($OS_I$), and K8s installation ($K8S_I$). As the cluster scales from 3 to 9 nodes, the end-to-end cluster deployment time increases from 532.3 s to 1023.7 s (+92.3%). The $K8S_I$ phase dominates across all configurations (52-55%), followed by $OS_I$ (30-31%) and *ND* (14-16%). In absolute terms, *ND* scales from 73.4 s to 168.8 s, $OS_I$ from 167.7 s to 319.9 s, and $K8S_I$ from 291.2 s to 535.0 s, confirming its primary scaling driver.

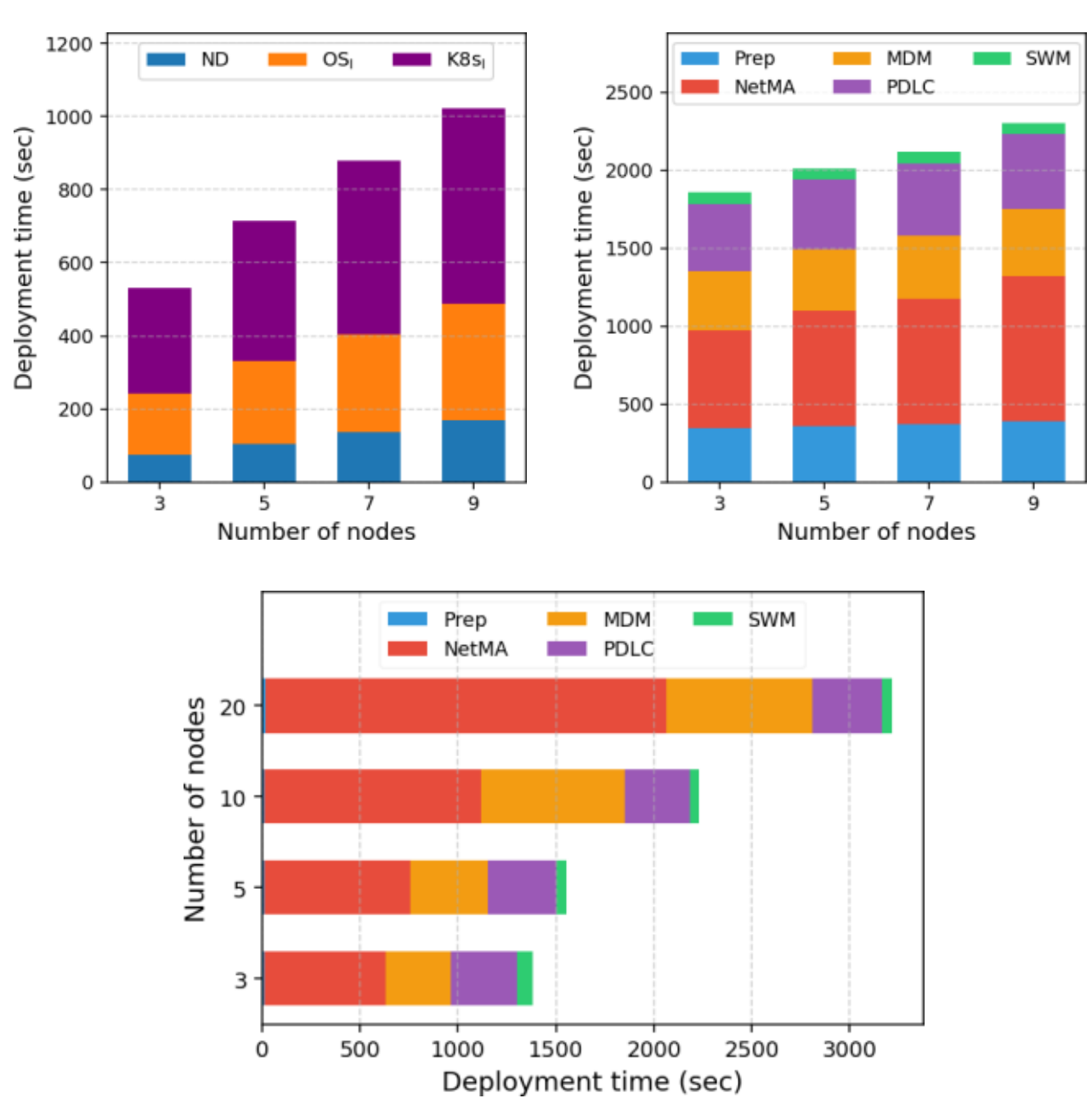

Figure 4: Mean deployment time per nodes' number: (a) CODEF stages (RD, OS & K8s installation), (b) CODECO installation via CODEF, and (c) CODECO installation via KinD.

Fig. 4(b) showcases CODECO installation times per component. For 3 nodes, installation requires 1855.9 s, increasing to 2011.1 s (5 nodes), 2117.7 s (7 nodes), and 2303.1 s for 9 nodes (+24.1% overall). NetMA constitutes the largest and fastest-growing component, rising from 624.6 s (33.7%) at 3 nodes to 932.4 s (40.5%) at 9 nodes, driven by its network configuration overhead scaling with cluster size. PDLC follows as the second largest, increasing from 429.0 s (23.1%) to 483.5 s (21.0%), while MDM grows moderately from 383.9 s (20.7%) to 427.3 s (18.5%). Preparation remains relatively stable, from 346.1 s (18.6%) to 389.2 s (16.9%), while SWM remains consistently minimal across all configurations, ranging from 72.3 s to 70.7 s (<4%).

Fig. 4(c) reports CODECO installation time overhead per component in a KinD environment (in INF1). For 3 nodes, the total CODECO installation requires 1380 s, increasing to 1554.0 s for 5 nodes (+12.6%), 2232.0 s for 10 nodes (+61.7%) and 3218.0 s for 20 nodes (+133.2% against 3 nodes). In terms of components, NetMA (network management) is the dominant time-consuming factor (44.5% at 3 nodes, reaching 63.6% at 20 nodes), followed by MDM (metadata management, 24.3–23.2%) and PDLC (intelligent decision making, 24.3–11.2%).

### 5.3. Pod / Service Startup and Deletion Times

This section evaluates pod startup and deletion latency for three applications of increasing complexity (pause pods, dummy frontend-backend, and Bookinfo) comparing CODECO against vanilla K8s across diverse infrastructure configurations and varying deployment scales.

#### 5.3.1. Application 1: Pod readiness per pod batch

Fig. 5 (top row) compares baseline K8s against CODECO in terms of pod startup latency for batch deployments of *pause* pods (10-150). Although CODECO adds orchestration logic beyond vanilla K8s, the resulting readiness overhead remains modest in absolute terms and represents a small cost for the higher-level scheduling and cross-layer capabilities CODECO provides.

Across the evaluated UCs, CODECO exhibits varying behavior relative to baseline K8s, with results depending on both batch size and underlying infrastructure. In particular, CODECO is consistently slightly slower than K8s, with the gap increasing with batch sizes but remaining within a few seconds, even at 150 pods. In INF2, mean readiness rises from 5.46 s (K8s: 3.98 s) at 10 pods to 57.11 s (K8s: 37.73 s) at 150 pods (+51.4%). In INF3, the overhead is 8.35 s (+15.4%) at 150 pods, while INF6 shows +5.71 s (+20.7%). INF4 presents an exception where CODECO matches or slightly outperforms K8s (e.g., 46.96 s against 47.96 s at 150 pods), suggesting that CODECO's scheduling optimizations can potentially counterbalance orchestration overhead under certain configurations. Finally, INF5 exhibits the highest readiness times across all batch sizes, reflecting the constrained compute capacity of RPi workers. At 10 pods, K8s reaches readiness in 7.5 s while CODECO requires 10.5 s (+40%), while at 100 pods the difference reaches 12%. The 150-pod batch was not deployed due to resource limitations.

Variability, as captured by standard deviation (*STDIV*), also differs across UCs and platforms. In INF2, K8s exhibits higher variability at 50 pods (1758 ms) compared to CODECO (1613 ms), whereas at 150 pods both platforms show reduced variance (K8s at 384 ms, while CODECO at 1282 ms). INF3 demonstrates low variability for K8s (41-1671 ms across scales) and moderate variability for CODECO (348-815 ms), while INF6 (AWS-based VMs) exhibits the most consistent behavior, with *STDIV* below 400 ms. For INF5, the STDIV ranges between 488–2105 ms for CODECO, consistent with the constrained but stable nature of the RPi hardware.

Readiness times differ substantially across UCs due to infrastructure heterogeneity. For example, at 50 pods under K8s, INF3 reaches 18.05 s whereas INF6 reaches 9.84 s, reflecting variations in hardware profiles. INF2 exhibits intermediate behavior (14.77 s at 50 pods), while INF4 demonstrates the fastest K8s performance (12.11 s at 50 pods). Finally, INF5 exhibits the slowest baseline readiness (20 s at 50 pods for K8s), as a result of the ARM-based RPi workers operating under k3s.

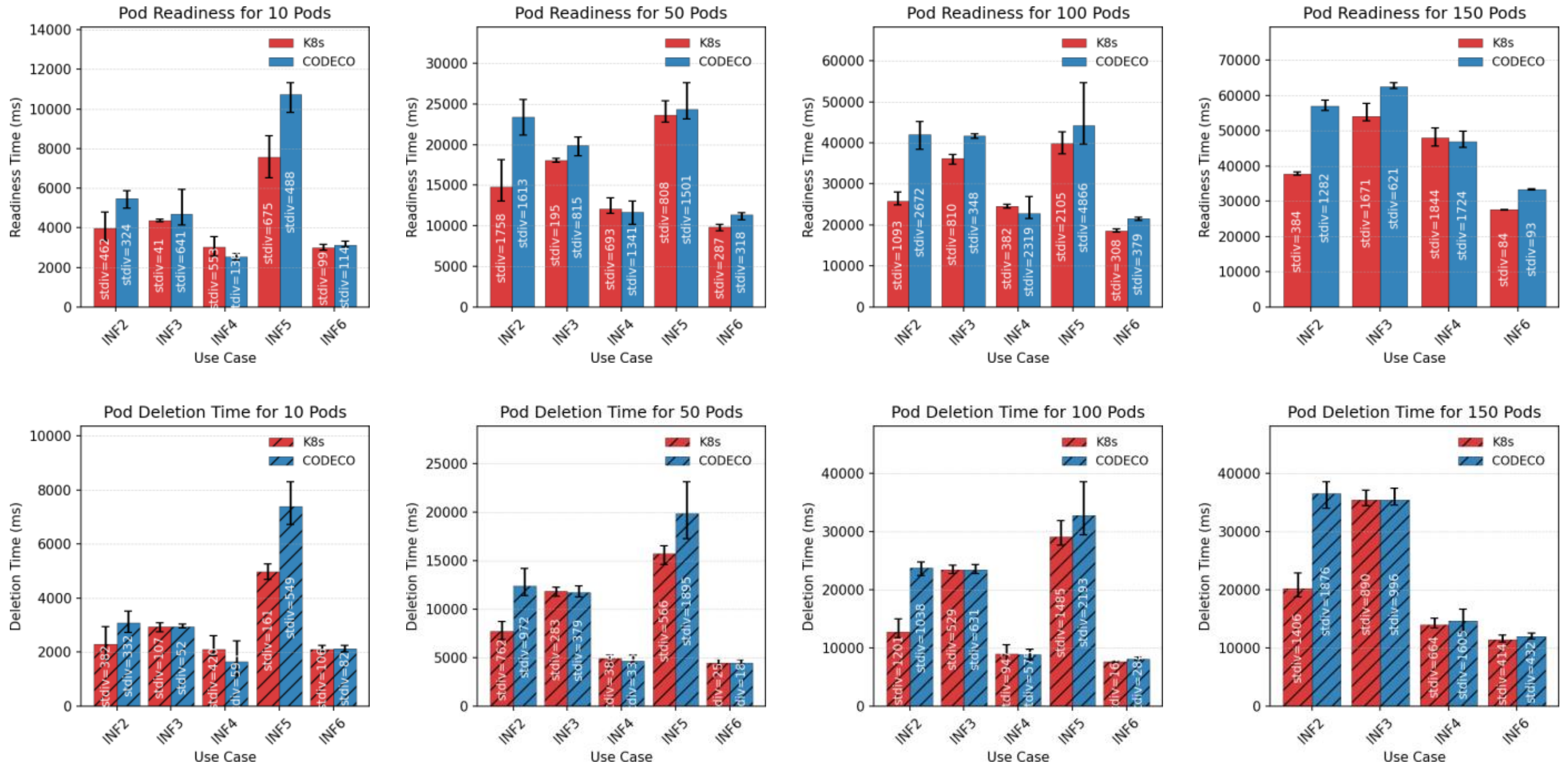

Figure 5: Mean pod readiness time (top row) and deletion time (bottom row) for a batch of pause pods (10–150).

In terms of pod deletion (Fig. 5 (bottom row)), CODECO maintains deletion performance comparable to vanilla K8s across all evaluated batch sizes (10-150 pods). In INF3 and INF6, CODECO's mean deletion times closely track those of K8s, indicating that CODECO's scheduling does not introduce significant teardown overhead. The general behavior between startup and deletion confirms that CODECO primarily impacts the pod initialization phases, where cross-layer scheduling decisions are made, rather than the termination sequence. INF5 constitutes an exception, exhibiting higher deletion latency than other UCs, which is consistent with the slower I/O and processing capabilities of RPi-based nodes.

### 5.3.2. Application 2: Dummy Frontend-Backend

Fig. 6 (top row) shows the mean readiness time of the dummy frontend-backend application while scaling the number of *frontend* replicas (1-50). Compared to the pause-pod experiment (rf. to Section 5.3.1), this captures service-level startup behavior, i.e., the impact of deploying a structured application (multiple replicas and K8s objects) under baseline K8s against CODECO.

The dummy application experiments also reveal substantial performance variation across use cases. In INF6, K8s scales from 4.22 s (1 replica) to 18.38 s (50 replicas), while CODECO increases from 5.67 s to 32.03 s, with overheads ranging from 1.45 s (+34.5%) to 13.65 s (+74.3%). In INF2, both vanilla K8s and CODECO exhibit higher times due to heavier virtualization overhead. K8s ranges from 5.44 s to 48.35 s, while CODECO fluctuates between 13.53-52.95 s. An interesting observation is the overhead CODECO narrows at scale, from +148.7% (1 replica) to +9.5% (50 replicas), suggesting that orchestration cost becomes proportionally smaller as workload complexity increases. INF5 follows a similar trend, with K8s scaling from 6 s (1 replica) to 36 s (50 replicas) and CODECO from 10 s to 33 s. Particularly, at higher replica counts, the overhead introduced by CODECO decreases, following a pattern similar to that observed in INF2.

Standard deviation analysis reveals infrastructure-dependent profiles. INF6 exhibits low variability for K8s (223-338 ms) but higher variance for CODECO, particularly at 50 replicas (4217 ms). INF4 shows high K8s variability at 1 replica (3.78-10.19 s), while CODECO maintains tighter distributions at 25 replicas (15.64-15.74 s). Finally, INF2 demonstrates moderate variability for both cases, with *STDIV* ranging from 370 ms to 4497 ms.

In terms of deletion, Fig. 6 (bottom row) reports the mean pod deletion times. In INF6, CODECO showcases a higher deletion latency than vanilla K8s, with an overhead in the order of tens of seconds. In contrast, K8s increases from 2.60 s (1 replica) to 7.47 s (50 replicas), while CODECO increases from 3.97 s to 18.92 s, corresponding to overheads of 1.37 s (1 replica), 5.89 s (10 replicas), 5.45 s (25 replicas) and 11.44 s (50 replicas). In INF5, deletion times for both K8s and CODECO are comparable across all replica counts, with CODECO occasionally matching or slightly outperforming K8s, indicating that the infrastructure does not amplify teardown overhead.

Note that for the 50-replica configuration, some UCs (e.g., INF3) are not included due to cluster resource constraints and configuration limitations, consequently, the corresponding metrics were excluded from the analysis.

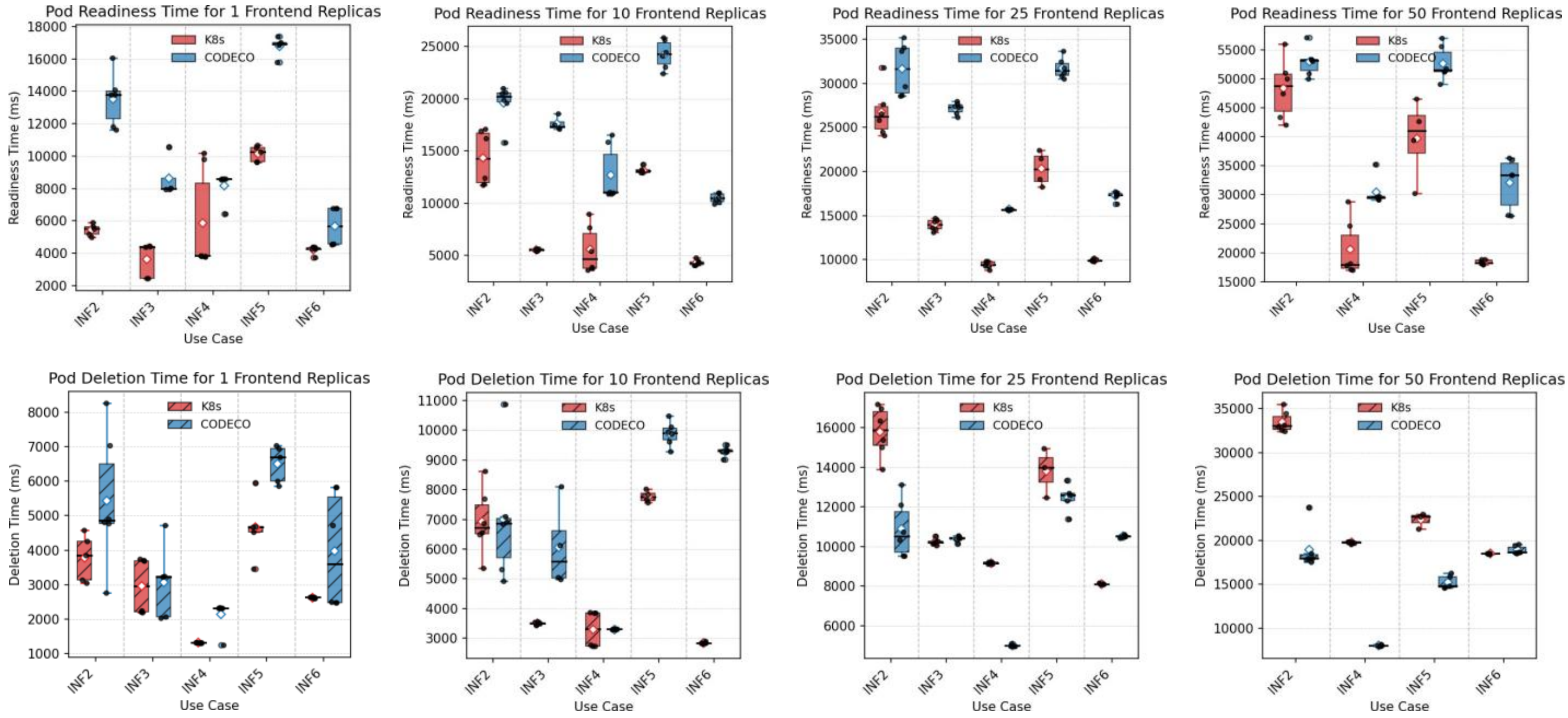

Figure 6: Mean pod readiness time (top row) and deletion time (bottom row) for CODECO dummy (frontend-backend) app with varying frontend replicas (1–50).

### 5.3.3. *Application 3: Bookinfo*

Fig. 7 showcases the mean deployment (readiness) and deletion times for the Bookinfo application under baseline K8s and CODECO. Across the evaluated use cases, CODECO introduces measurable startup overhead that varies by infrastructure. In INF6, K8s reaches readiness in 3.5 s on average, while CODECO requires 11 s. INF3 exhibits similar behavior with K8s completed in 3 s against CODECO at 13 s, while INF2 shows K8s at 4.5 s and CODECO at 14 s. INF4 presents the smallest gap, with K8s at 6.5 s and CODECO at 9 s. Finally, INF5 exhibits the largest readiness times among all UCs, with K8s at approximately 20 s and CODECO at 30 s, reflecting the resource constraints of RPi-based edge nodes.

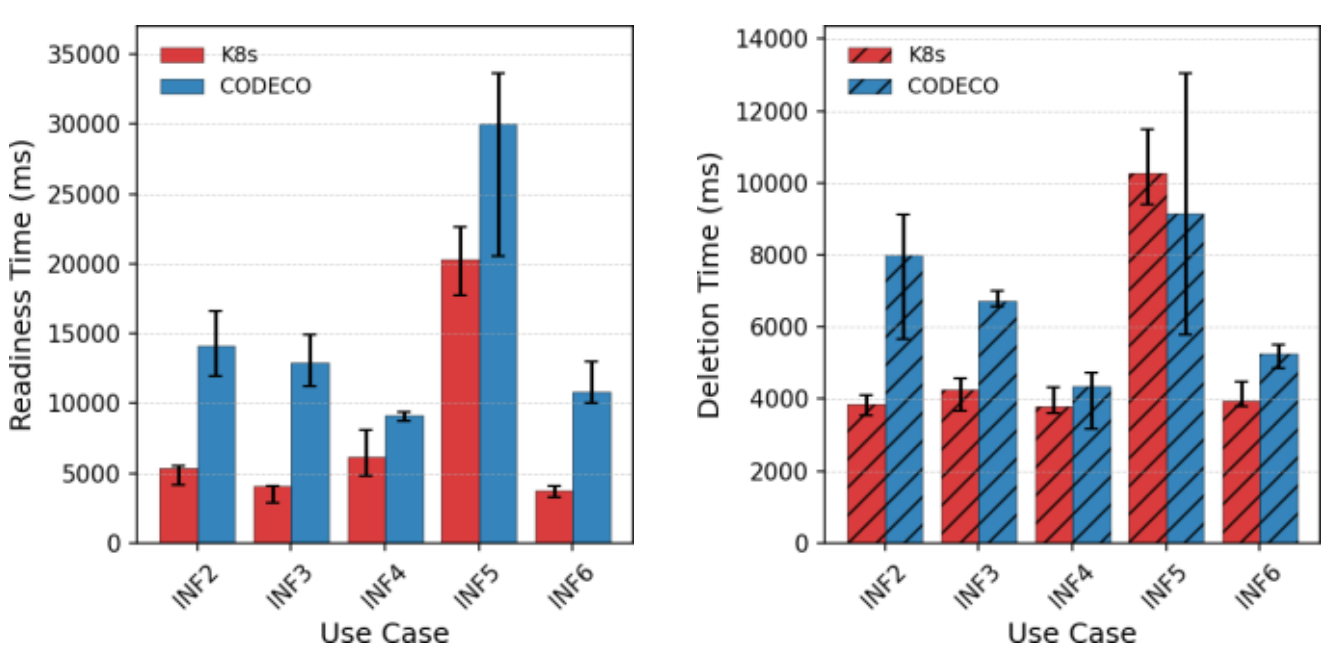

Figure 7: Mean deployment (left) and deletion times (right) for the Bookinfo app in vanilla K8s and CODECO environments.

For deletion, CODECO's overhead across all UCs ranges from +0.60 s in INF4 (+16.2%) to +4.30 s in INF2 (+116.2%). K8s completes teardown in 3.70-4.20 s, while CODECO requires 4.30-8 s. Regarding variability, K8s maintains stable performance (261-415 ms), while CODECO ranges from 165 ms for INF3 and INF6, to higher variance in INF2 (2388 ms). Similarly to readiness results, INF5 exhibits the highest baseline deletion time (10.5 s for K8s and 9.5 s for CODECO).

Despite the overhead compared to previous experiments and applications, the absolute times remain below 15 s for startup and below 10 s for deletion, which is acceptable for Edge-Cloud deployments, especially given CODECO's cross-layer advanced capabilities.

### 5.4. *Resource Footprint Metrics*

Fig. 8 and Table 4 report the resource footprint (CPU, memory and energy) for master and worker nodes, as their number increases. The clusters are deployed with CODEF, comparing K8s against CODECO. Once the cluster reached a stable state, metrics were gathered over a 30-minute period, with the final values representing the mean performance across this interval (rf. to Table 4).

In idle state, vanilla K8s exhibits minimal CPU overhead with the master node having an average of 1.76%, while workers consume 1.10% and 0.69% respectively. CODECO introduces additional CPU load due to its orchestration components. The master node reports 4.66% (+2.90%), while workers reach 3.81% and 2.30% (+2.71% and +1.61%). Cluster-wide, the total idle CPU rises from 3.55% (K8s) to 10.78% (CODECO). CODECO reports higher variability, reaching 15% on workers, compared to K8s's stable range of 0.58-2.26%.

The memory overhead is more notable as CODECO significantly expands the cluster's resource requirements. While vanilla K8s deployment maintains a footprint for the master at 1.15 GB and workers at 0.99 GB and 0.57 GB, CODECO increases memory consumption. The master node rises to 3.14 GB (+1.99 GB), while the workers reach 2.92 GB and 1.22 GB, respectively. It is important to note, however, that the absolute numbers are also influenced by the additional overhead introduced by the CODEF.

Table 4: Average resource footprint and estimated energy consumption for vanilla K8s and CODECO deployments in INF6 (3-node cluster in idle state).

| | CPU (%) | | Mem (GB) | | Power (W) | |
|---|---|---|---|---|---|---|
| **Node** | K8s | CODECO | K8s | CODECO | K8s | CODECO |
| Master | 1.76 | 4.66 | 1.15 | 3.14 | 36.3 | 38.3 |
| Worker 1 | 1.10 | 3.81 | 0.99 | 2.92 | 35.8 | 37.8 |
| Worker 2 | 0.69 | 2.30 | 0.57 | 1.22 | 35.5 | 36.6 |
| **Total** | 3.55 | 10.78 | 2.71 | 7.28 | 107.6 | 112.7 |
| **Overhead** | +7.23% | | +4.57 | | +5.1 (+4.7%) | |
| **Daily Energy (kWh)** | | | | | 2.58 | 2.70 |

**Power estimation parameters (Intel 8259CL): $P_{idle}$ = 35W, $P_{max}$ = 95W* per node

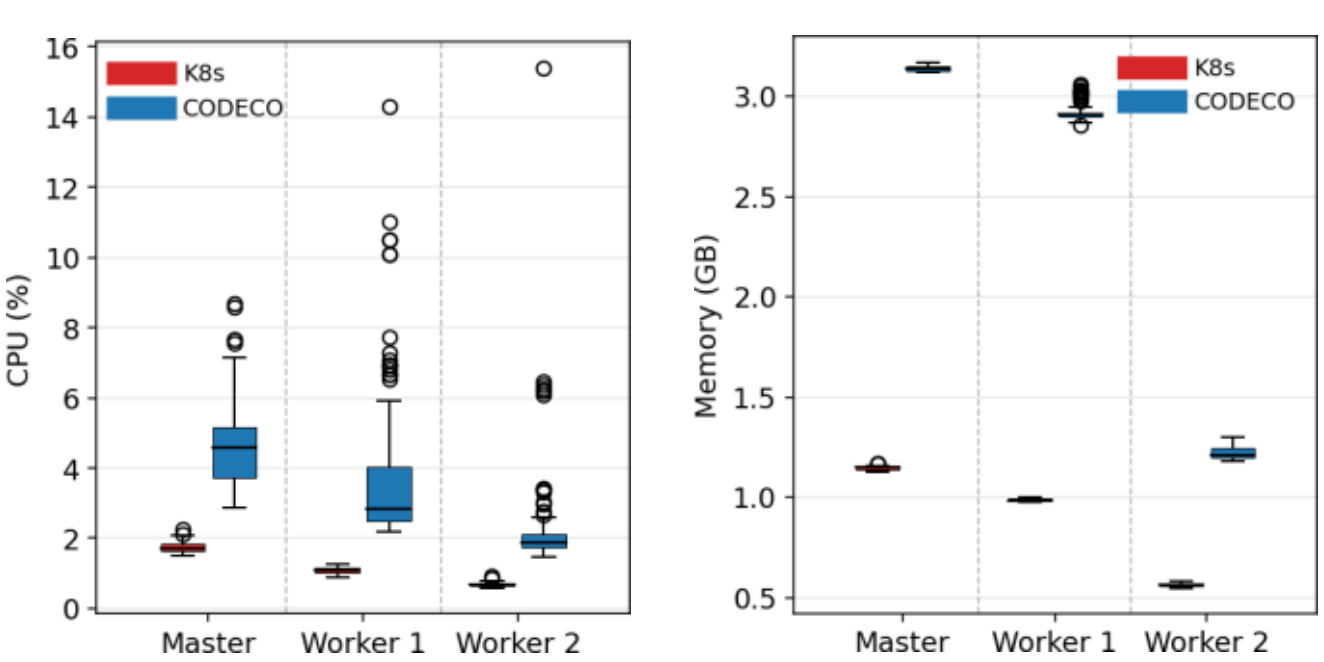

Figure 8: Average (i) CPU utilisation (%) and (ii) memory usage (GB) of master and worker nodes for vanilla K8s and CODECO in idle state with CODEF for INF6.

In terms of energy consumption, we estimate instantaneous power draw using the established linear power model $P = P_{idle} + (P_{max} - P_{idle}) \times U_{cpu} + P_{mem}$, with parameters derived from INF6 specifications (rf. to Tables 2). Under idle conditions, the vanilla K8s cluster consumes approximately 107.6 W across all three nodes, i.e., 2.58 kWh/day, while CODECO increases the total power consumption to 112.7 W (2.70 kWh/day), a 4.7% energy overhead. Table 4 summarizes the per-node breakdown and estimated energy.

Similarly, Fig. 9 illustrates the system-level resource footprint (CPU, memory) for KinD deployments, comparing KinD against KinD + CODECO as the cluster scales from 3 to 20 nodes. Unlike CODEF, where each node runs on a dedicated VM, KinD deploys all K8s nodes as containers on a single host, resulting in aggregated resource consumption.

In idle state, KinD presents low CPU utilization, ranging from 0.6% (3 nodes) to 3.1% (20 nodes). CODECO increases this consumption to 2.3%-7.5%. Memory footprint follows a similar trend with KinD consuming 4.91-6.28 GB, while CODECO requires 12.18-17.70 GB, representing an overhead of +7.27 GB (3 nodes) to +11.42 GB (20 nodes). In terms of energy (rf. to Table 5), at 20 nodes, KinD consumes 64.5 W (1.55 kWh/day), while CODECO increases this to 71.4 W (1.71 kWh/day), a 10.7% overhead.

Table 5: Average resource footprint and estimated energy consumption for KinD and KinD + CODECO deployments in INF1 (scaling nodes, idle state).

| | CPU (%) | | Mem (GB) | | Power (W) | |
|---|---|---|---|---|---|---|
| **Nodes** | KinD | CODECO | KinD | CODECO | KinD | CODECO |
| 3 | 0.6 | 2.3 | 4.91 | 12.18 | 61.6 | 64.9 |
| 5 | 1.8 | 2.9 | 4.98 | 12.92 | 62.9 | 65.6 |
| 10 | 2.4 | 4.2 | 5.42 | 14.88 | 63.6 | 67.4 |
| 20 | 3.1 | 7.5 | 6.28 | 17.70 | 64.5 | 71.4 |
| **Overhead** | +4.4% | | +11.42 | | +6.9 (+10.7%) | |
| **Daily Energy (kWh)** | | | | | 1.55 | 1.71 |

**Power estimation parameters (i9-10900K): $P_{idle}$ = 60W, $P_{max}$ = 165W*

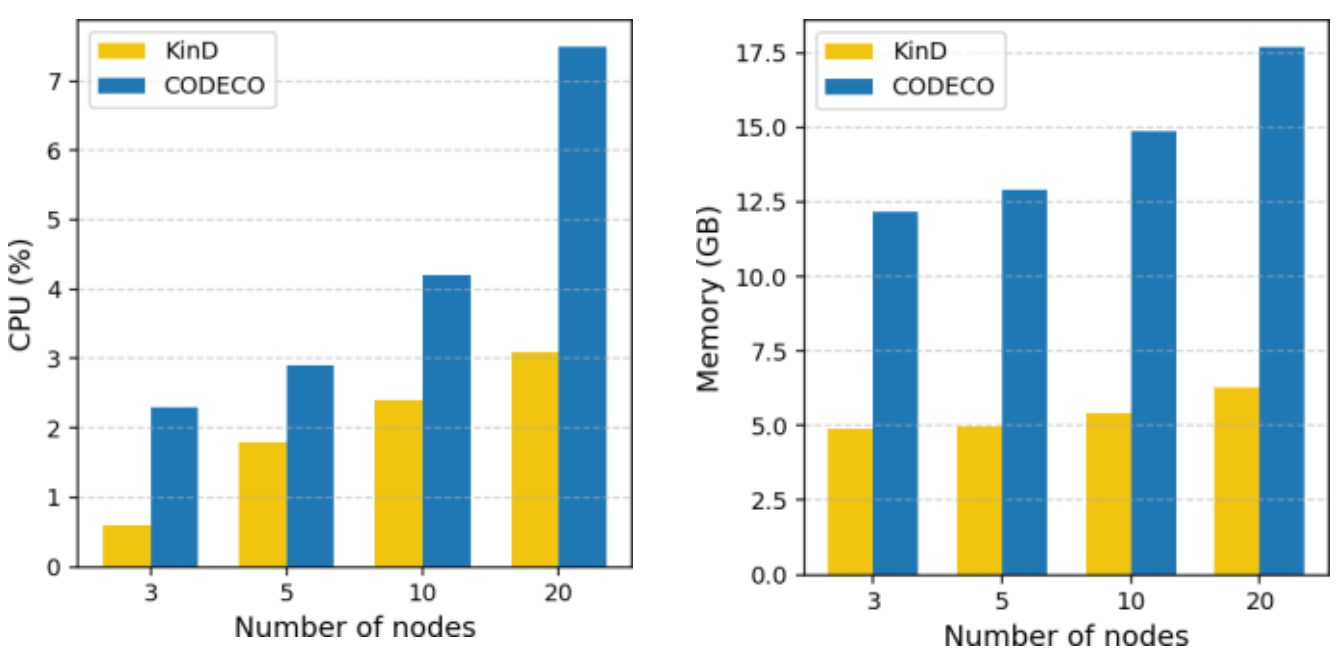

Figure 9: Average system (i) CPU utilisation (%) and (ii) memory usage (GB) for KinD and KinD + CODECO deployments in idle state for INF1.

## 6. Conclusion and Future Work

This paper presented an experimental evaluation of the CODECO OSS toolkit for multi-service deployment in single-cluster Edge-Cloud environments. CODECO was benchmarked against baseline K8s across heterogeneous infrastructures, workflows and K8s distributions. The study focused on three KPIs: (i) time taken to set up and configure clusters and services (deployment and lifecycle latency), (ii) percentage of manual intervention required during setup and deployment (via the MIP metric), and (iii) overall resource footprint in terms of CPU, memory and energy consumption. Our key observations are listed below:

- **Substantial automation gains**: CODEF and CODECO reduce manual intervention by 74.7% for cluster deployment, 90.5% for K8s installation, and 81.8% for service deployment compared to baseline K8s.

- **Component-level insights**: NetMA emerges as the dominant time-consuming component during CODECO installation (34–64% depending on cluster size and environment), indicating that network management orchestration represents the primary scaling challenge in distributed Edge–Cloud deployments.

- **Application-dependent overhead characteristics**: CODECO's orchestration overhead exhibits diverse behavior across application complexity levels. For minimal batch pods, overhead remains acceptable (within seconds even at 150 pods), while more realistic microservices show variable patterns. In INF2 the relative overhead decreases at scale (from +148.7% at 1 replica to +9.5% at 50 replicas for the dummy app), suggesting that CODECO's orchestration cost becomes proportionally smaller as workload complexity increases.

- **Acceptable CPU and energy footprint with high memory overhead**: In idle state, CODECO adds +7.23% CPU overhead, +4.57 GB memory consumption and +4.7% energy overhead for CODEF deployments, while for KinD deployments +4.4% CPU, +11.42 GB memory and +10.7% energy (20 nodes), demonstrating the introduced resource overhead, primarily represented in memory usage increase.

- **Consistent operation across heterogeneous edge platforms** (RPi-based nodes): Experiments on constrained edge hardware with k3s validate CODECO across diverse edge configurations, despite performance variation (e.g., at 150 pause pods, readiness ranges (from 33.3 s on INF6 to 62.5 s on INF4) driven by hardware profiles. High-replica deployments exposed resource limitations on certain edge clusters (e.g., 150 pods for INF3), highlighting the importance of resource-aware scheduling, a capability CODECO natively supports. Particularly, INF5 (RPi-based cluster) consistently exhibited the highest latency across all experiments, yet CODECO maintained stable operation and, in some cases, achieved teardown times comparable to or lower than baseline K8s.

In terms of future work, we will extend our evaluation along several directions. First, we plan to move from single-cluster deployments to federated multi-cluster Edge-Cloud environments, assessing CODECO's orchestration capabilities across geographically distributed sites under variable inter-site connectivity, latency constraints and resource availability patterns. Finally, we will extend both metrics and experimental scenarios by incorporating richer workload dynamics (bursty traffic and mixed batch/interactive workloads) and fault injection conditions (node failures, network partitions or resource exhaustion).